\documentclass[prd,showpacs,superscriptaddress]{revtex4}

\usepackage{amsfonts}
\usepackage{amsmath}
\usepackage{amssymb}
\usepackage{amsthm}
\usepackage{bm}
\usepackage{dcolumn}
\usepackage{epsfig}
\usepackage{graphicx}
\usepackage{graphics}
\usepackage[latin1]{inputenc}
\usepackage{latexsym}
\usepackage{rotating}
\usepackage{hyperref}
\usepackage{url}

\newcommand{\beqs}{\begin{equation*}}
\newcommand{\beq}{\begin{equation}}

\newcommand{\eeqs}{\end{equation*}}
\newcommand{\eeq}{\end{equation}}

\newcommand{\beqas}{\begin{eqnarray*}}
\newcommand{\beqa}{\begin{eqnarray}}

\newcommand{\eeqas}{\end{eqnarray*}}
\newcommand{\eeqa}{\end{eqnarray}}

\newcommand{\eq}[2]{\begin{equation} #1 \label{#2} \end{equation}}

\newcommand{\eqa}[2]{\begin{eqnarray} #1 \label{#2} \end{eqnarray}}

\newcommand{\de}{\delta}
\newcommand{\om}{\omega}

\newcommand{\De}{\Delta}

\newcommand{\La}{\Lambda}

\newcommand{\blist}{\begin{itemize}}

\newcommand{\elist}{\end{itemize}}

\providecommand{\href}[2]{#2}

\DeclareFontFamily{OT1}{rsfs}{}
\DeclareFontShape{OT1}{rsfs}{m}{n}{ <-7> rsfs5 <7-10> rsfs7 <10->rsfs10}{} 
\DeclareMathAlphabet{\mycal}{OT1}{rsfs}{m}{n}

\DeclareMathOperator{\extdm}{d}
\newcommand{\extd}{\extdm \!}

\begin{document}
\title{Solar system constraints on Rindler acceleration}

\author{Sante Carloni}
\affiliation{European Space Research Technology Center, Space Science Department, Keplerlaan 1, Postbus 299, 2200 AG Noordwijk, The Netherlands}
\author{Daniel Grumiller}
\affiliation{Institute for Theoretical Physics, Vienna University of Technology, Wiedner Hauptstr. 8-10/136, A-1040 Vienna, Austria}
\author{Florian Preis}
\affiliation{Institute for Theoretical Physics, Vienna University of Technology, Wiedner Hauptstr. 8-10/136, A-1040 Vienna, Austria}

\date{\today}

\preprint{TUW-11-xx}

\begin{abstract}
We discuss the classical tests of general relativity in the presence
of Rindler acceleration. Among these tests the perihelion shifts give
the tightest constraints and indicate that the Pioneer
anomaly cannot be caused by a universal solar system Rindler
acceleration. We address potential caveats for massive test-objects. Our tightest bound on Rindler acceleration that comes with no caveats is derived from radar echo delay and yields $|a|<3nm/s^2$.
\end{abstract}


\maketitle


\section{Introduction}

Recently one of us proposed a model for gravity at large distances \cite{Grumiller:2010bz}, which was derived from postulating spherical symmetry, diffeomorphism invariance as well as additional technical requirements.
These assumptions led to the line-element
\begin{align}
& 
\extd s^2 = -K^2\extd t^2+\frac{\extd r^2}{K^2} + r^2\,\big(\extd\theta^2+\sin^2\!\theta\extd\phi^2\big)
\label{eq:rind8} 
\\
& 
K^2 = 1-\frac{2M}{r}-\Lambda r^2+2ar
\label{eq:rind9}
\end{align}
which reduces to the Schwarzschild line-element (with mass $M$) for vanishing Rindler acceleration $a=0$ and vanishing cosmological constant $\La=0$.
Thus, in comparison to spherically symmetric vacuum general relativity a novel term in the Killing norm $K$ arises proportional to a new constant $a$.  
The effective model constructed in \cite{Grumiller:2010bz} does not make any prediction about the magnitude or sign of $a$.
Neither does it make a clear statement whether or not $a$ is universal (like the cosmological constant $\Lambda$) or system dependent (like the mass $M$).
It is of interest to constrain $a$ from observations of/for relevant astrophysical systems.

The main purpose of the current paper is to study the classical solar system tests of general relativity --- perihelion shifts, light bending and gravitational redshift --- in the presence of Rindler acceleration $a\neq 0$. 
We set $\La=0$ since the cosmological constant does not influence the solar system dynamics appreciably; $M$ is the solar mass and $a$ is assumed to be some constant the value of which we would like to constrain. 
To this end we study the motion of test particles in the background \eqref{eq:rind8}, \eqref{eq:rind9} (with $\La=0$).
Using standard methods \cite{waldgeneral} yields the well-known result $\dot\phi=\ell/r^2$, where $\ell$ is the conserved angular momentum, and
\eq{
\frac{\dot r^2}{2}+ V^{\rm eff} = E
}{eq:rind10}
with $E=\rm const.$
For time like test particles the effective potential reads \cite{Grumiller:2010bz}
\eq{
V^{\rm eff} = -\frac{M}{r} + \frac{\ell^2}{2r^2} - \frac{M\ell^2}{r^3} + ar\,\big(1+\frac{\ell^2}{r^2}\big)\ .
}{eq:rind11}
For light like test particles the effective potential simplifies to
\eq{
V^{\rm eff} = \frac{\ell^2}{2r^2} - \frac{M\ell^2}{r^3} + \frac{a\ell^2}{r}\ .
}{eq:ct1}
The potentials above will play a crucial role in our discussion of the classical tests.
The observational data will then establish upper bounds on the magnitude of Rindler acceleration.

A simple bound can be deduced directly from the Pioneer data \cite{Anderson:1998jd}.
Namely, if the Rindler constant $a$ was bigger than the anomalous Pioneer acceleration $a_{\rm Pioneer}$ it should have shown up already in the data of the Pioneer trajectory.
This statement is true regardless of whether the ``Pioneer anomaly'' is a physical effect or an artifact.
Thus, we obtain an upper bound on $a$:
\eq{
|a| \leq a_{\rm Pioneer} = 1.6\cdot 10^{-61} \sim 9 \cdot 10^{-10} m/s^2
}{eq:Pioneerbound}
In the remainder of the paper we investigate to what extent the classical tests of general relativity can make the bound \eqref{eq:Pioneerbound} stronger.

This paper is organized as follows.
In section \ref{se:2} we derive the perihelion shift and compare with the observational data to obtain upper bounds on the Rindler acceleration.
In section \ref{se:3} we discuss light bending and radar echo delay.
In section \ref{se:4} we show that gravitational redshifts do not pose strong constraints on Rindler acceleration and also address further tests.
We summarize and discuss our results in section \ref{se:5}.

Before starting let us mention some of our conventions.
We use signature $(-,+,+,+)$ and natural units $G_N=\hbar=c=1$, unless stated otherwise.

\section{Perihelion shift}\label{se:2}

We derive now a result for perihelion shift in the limit of small eccentricity, small Rindler acceleration and small general relativistic corrections, following Ref.~\cite{waldgeneral}.
Namely, if a particle is perturbed slightly around some stable circular orbit at radius $r=r_+$ then it will oscillate harmonically with frequency given by
\eq{
\om_r^2 = \frac{\extd^2 V^{\rm eff}}{\extd r^2}\Big|_{r=r_+} = -\frac{2M}{r_+^3} + \frac{3\ell^2}{r_+^4}-\frac{12M\ell^2}{r_+^5}+\frac{2a\ell^2}{r_+^3}\ .
}{eq:ct2}
The angular momentum is determined from the condition $\extd V^{\rm eff}/\extd r|_{r=r_+} = 0$.
\eq{
\ell^2 = Mr_+ \Big( 1 + \frac{3M}{r_+} + \frac{ar_+^2}{M} + 2 ar_+\Big) + \dots
}{eq:ct3}
The ellipsis denotes higher order terms that we neglect.
The perihelion precession is given by the difference between the angular frequency $\om_\phi=|\ell|/r_+^2$ and the frequency $\om_r$ \eqref{eq:ct2}.
\eq{
\om_p = \om_\phi-\om_r = \frac{3M^{3/2}}{r_+^{5/2}}\,\Big(1-\frac{ar_+^3}{3M^2}\Big) + \dots
}{eq:ct4}
The higher order corrections include finite eccentricity corrections, terms of higher order in $ar_+$ and terms of higher order in $M/r_+$. We note that the leading Rindler correction in the perihelion precession \eqref{eq:ct4} does not originate from the term proportional to the Rindler acceleration in the radial frequency \eqref{eq:ct2}, but rather from the leading Rindler correction to the angular momentum \eqref{eq:ct3}.

A slightly more elaborate analysis analogous to Ref.~\cite{Weinberg:1972} establishes a result that is valid also for finite eccentricities:
\eq{
\om_p = \frac{3M^{3/2}}{(1-e^2)A^{5/2}}\,\Big(1-\frac{aA^3(1-e^2)^{3/2}}{3M^2}\Big) + \dots
}{eq:ct5}
Here $A$ is the semimajor axis of the ellipse and $e$ its eccentricity.
The leading order term in front of the parenthesis in \eqref{eq:ct5} is the general relativistic result.
The leading order correction in the parentheses, $aA^3(1-e^2)^{3/2}/3M^2$, therefore must be sufficiently small in order not to conflict with observational data.
Note that this correction diminishes the perihelion shift for positive Rindler constant $a>0$.
All omitted terms in \eqref{eq:ct5} are subleading.
Requiring the contribution $aA^3(1-e^2)^{3/2}/3M^2$ to be small establishes a constraint on the magnitude of the Rindler acceleration $|a|$.
\begin{table}
\begin{tabular}{||c||c|c|c|c|c|c|c|c|c||} \hline\hline
Planet & Mercury & Venus & Earth & Mars & Jupiter & Saturn & Uranus & Neptune & Icarus  \\\hline
$A$ & $4\cdot 10^{45}$ & $7\cdot 10^{45}$  & $9 \cdot 10^{45}$ & $1.4\cdot 10^{46}$ & $5\cdot 10^{46}$ & $9\cdot 10^{46}$ & $1.8\cdot10^{47}$ & $3\cdot 10^{47}$ & $1.0\cdot10^{46}$ \\ 
$e$ & 0.2 & 0.007 & 0.017 & 0.09 & 0.05 & 0.06 & 0.05 & 0.011 & 0.8 \\ 
$\De\phi=2\pi\om_p A^{3/2}/\sqrt{M}$ & 43 & 8.6 & 3.8 & 1.3 & 0.06 & 0.014 & 0.002 & 0.0007 & 9.8 \\
$\de\om_p/\om_p$ & $1.2\cdot 10^{-4}$ & $3\cdot 10^{-2}$ & $1.1\cdot 10^{-4}$ & $4\cdot 10^{-4}$ & $0.6$ & $3\cdot 10^{2}$ & $7\cdot 10^3$ & ? & $8\cdot10^{-2}$ \\ \hline
$|a|<$ & $7\cdot 10^{-65}$ & $3\cdot 10^{-63}$ & $4\cdot 10^{-66}$ & $4\cdot 10^{-66}$ & $1.2\cdot 10^{-64}$ & $7\cdot 10^{-63}$ & $3\cdot 10^{-62}$ & ? & $1.1\cdot 10^{-62}$ \\ \hline\hline
\end{tabular}
 \caption{Semimajor axes $A$, eccentricities $e$, perihelion shifts $\De\phi$ (in arcseconds per century), residual relative perihelion shifts $\de\om_p/\om_p$ and constraints on the Rindler acceleration $|a|$ for all solar system planets}
 \label{tab:1}
\end{table}

We address now the observational data that we collect in table \ref{tab:1}.
The data for the semimajor axes and eccentricities are well-known, while the data for the perihelion shifts~\footnote{When referring to perihelion shifts we always mean the amount of perihelion shift not explained by Newtonian gravity, i.e., the perihelion shift described by general relativity and physics beyond general relativity.} and the residual perihelion shifts of the planets Mercury, Venus, Earth and Mars were taken from Ref.~\cite{Pitjeva:2005}. The perihelion shifts of the planets Jupiter, Saturn and Uranus were taken from general relativistic calculations. The residual perihelion shifts of the planets Jupiter, Saturn and Uranus were taken from Ref.~\cite{Iorio:2007ub} and the data for (1566) Icarus can be found in Ref.~\cite{Weinberg:1972}.
With the solar mass given by $M\approx 9\cdot 10^{37}$ the tightest constraints on solar Rindler acceleration come from the perihelion shifts of Earth and Mars:
\eq{
|a|<4\cdot 10^{-66} \sim 2\cdot 10^{-14} m/s^2
}{eq:ct6}
We postpone a discussion of this bound (and all further bounds found below) to section \ref{se:5}.
The order of magnitude of the bound \eqref{eq:ct6} can also be obtained from an analysis analogous to the one by Sereno and Jetzer \cite{Sereno:2006mw}, which constrains modifications of the gravitational inverse-square law of Yukawa-type or modified Newton dynamics \cite{Milgrom:1983ca}.

\section{Light bending and radar echo delay}\label{se:3}

In this section we consider the classical light bending and radar echo delay, following Ref.~\cite{Weinberg:1972}.
At the point where the photon's worldline is closest to the Sun, $r=r_0$, the quantity $\dot r$ in \eqref{eq:rind10} vanishes.
Thus, the effective potential \eqref{eq:ct1} evaluated at $r_0$ equals the energy $E$.
Plugging this value for $E$ into \eqref{eq:rind10} along the trajectory of the photon we then obtain 
\eq{
\frac{\extd r}{\extd\phi} = r^2K(r)\sqrt{\frac{K^2(r_0)}{r_0^2 K^2(r)}-\frac{1}{r^2}}
}{eq:ct7}
with $K$ defined in \eqref{eq:rind9} (again, we set $\La=0$).
With the substitution $y=r_0/r$, the deflection angle of a photon emitted and absorbed at $r\to\infty$ is given by
\eq{
\De\phi=2\int\limits_0^1 \frac{\extd y}{\sqrt{K^2(1)-K^2(y)y^2}}-\pi=\frac{4M}{r_0}\,\Big(1-\frac{ar_0^2}{2M}\Big)+\dots = \frac{4M}{r_0} + \de\De\Phi\ .
}{eq:ct8}
The leading order term in front of the parenthesis in \eqref{eq:ct8}, $4M/r_0$, is the general relativistic result.
The leading order correction in the parentheses, $ar_0^2/2M$, therefore must be sufficiently small in order not to conflict with observational data.
Note that this correction diminishes the light deflection for positive Rindler constant $a>0$.
All omitted terms in \eqref{eq:ct8} are subleading.

We address now the observational data.
From Ref.~\cite{Shapiro:2003} we obtain a bound on the residual deflection angle 
\eq{
-1.2\cdot10^{-3}<\frac{\de\Delta\phi}{\Delta\phi} < 4\cdot10^{-4}
}{eq:ct9}
extracted from their error bars for the PPN parameter $\gamma$. There the light bending data of quasars collected from 1979 to 1999 was used.
Taking for $r_0$ approximately twice the solar radius, $r_0\approx 8\cdot 10^{43}$ and inserting into our result \eqref{eq:ct8} the tighter of the two bounds in \eqref{eq:ct9} yields a very loose bound on Rindler acceleration.
\eq{
|a|< 1.1\cdot 10^{-53}\sim 6\cdot 10^{-2} m/s^2
}{eq:ct10}
Clearly, the bounds imposed by the perihelion shifts in the previous section are much tighter.

We consider next radar echo delay. 
Namely, we calculate the (coordinate) time delay due to light bending and clock effects for a radar signal sent from Earth to some planet or space craft and reflected back to Earth when Earth and the target are in opposition, see also Ref.~\cite{Weinberg:1972}.
We start from the differential equation
\eq{
\frac{\extd r}{\extd t}=K^2(r)\sqrt{1-\frac{K^2(r)r_0^2}{K^2(r_0)r^2}}\ .
}{eq:ct11a}
From that we compute the time delay
\eqa{
\De t &&=2\left(\int\limits_{r_0}^{r_E}\frac{\extd r}{K^2(r)}\left(1-\frac{K^2(r)r_0^2}{K^2(r_0)r^2}\right)^{-1/2}+\int\limits_{r_0}^{r_T}\frac{\extd r}{K^2(r)}\left(1-\frac{K^2(r)r_0^2}{K^2(r_0)r^2}\right)^{-1/2}-\sqrt{r_E^2-r_0^2}-\sqrt{r_T^2-r_0^2}\right)=\nonumber
\\&&=4M \Big(\ln{\frac{4r_E r_T}{r_0^2}}+1\Big) - 2a (r^2_E + r^2_T) + \dots = 4M \Big(\ln{\frac{4r_E r_T}{r_0^2}}+1\Big) + \de\De t\ .
}{eq:ct11}
Here $r_0$ is again a distance of the order of the solar radius, while $r_E$ and $r_T$ are the semimajor axes of Earth and the target orbits, respectively (see table \ref{tab:1}).
The first term on the second line of \eqref{eq:ct11} is the general relativistic result, while the second term is the leading correction from Rindler acceleration.
As usual, we have neglected subleading terms.

We address now the observational data.
From Ref.~\cite{Bertotti:2003} we obtain a bound on the residual time delay 
\eq{
-10^{-6}<\frac{\de\Delta t}{\Delta t}< 2\cdot 10^{-5}
}{eq:ct12} where the signal from the Cassini space craft was used.
Our result \eqref{eq:ct11} then yields a considerably tighter bound on Rindler acceleration compared to the one from light bending,
\eq{
|a|<5\cdot 10^{-61}\sim 3\cdot 10^{-9} m/s^2
}{eq:ct13}
with the parameters $r_0\approx 7\cdot 10^{43}$ and $r_T\approx 7\cdot 10^{46}$ also taken from Ref.~\cite{Bertotti:2003}.
The bound \eqref{eq:ct13} is close to (but slightly weaker than) the one obtained from the Pioneer acceleration \eqref{eq:Pioneerbound}.
Again, the bounds imposed by the perihelion shifts in the previous section are much tighter.

\section{Gravitational redshift and further tests}\label{se:4}

Gravitational redshift is traditionally included as a test of general relativity, but it is in fact a test of the (Einstein) equivalence principle \cite{will2010confrontation}. 
In our case, however, we can use it to 
restrict the Rindler term although, as we shall 
see, the precision of the data does not allow to put a strong constraint on the value of $a$.  The formula for the gravitational redshift of a light ray emitted in $E$ and received in $R$ by observers following the timelike Killing field in a general stationary metric $g_{\mu\nu}$ following Refs.~\cite{Weinberg:1972,waldgeneral} reads
\begin{equation}
\frac{\nu_E}{\nu_R}=\sqrt{\frac{g_{00}(R)}{g_{00}(E)}}\ .
\end{equation}
In the case of the metric (\ref{eq:rind9}) and upon assuming $\Lambda=0$, $ar_{I}\ll 1$, and $M\ll r_{I}$ with $I=R,E$ this formula takes the form
\begin{equation}\label{RedshiftNewt}
\frac{\nu_E}{\nu_R}=\frac{K(r_R)}{K(r_E)}\approx  1-M\left(\frac{1}{r_R}-\frac{1}{r_E}\right)+a \left(r_R-r_E\right)\ .
\end{equation}
This means that the Rindler correction increases the gravitational redshift effect by a factor that grows proportionally with the distance. 

At present there are many different measurements of the gravitational redshift which can be divided in two classes: the absolute redshift measurements and the null redshift experiments (see e.g. \cite{pound1960apparent,hafele1972around,hafele1972-obs,snider1972new,Colella:1975dq,vessot1979test,vessot1980test,Turneaure:1983zz,Bonse:1984td,Vucetich:1988uw,Horvath:1988vf,Krisher:1990zz,krisher1993galileo,godone1995null,Littrell:1997zz,Cottam:2002cu,cacciapuoti2009space}). Both of these classes of experiments are based on a simple Newtonian derivation of the deviation of the gravitational redshift due to Haugan \cite{Haugan:1979iv}. In  Ref.~\cite{Haugan:1979iv} the  gravitational redshift was expressed in terms of the difference $\Delta U$ in the (Newtonian) gravitational potential between the emission and the reception and a new parameter $\xi$ representing the deviation from this law was introduced:
\begin{equation}\label{z-original}
z=\frac{\nu_E}{\nu_R}-1=(1+\xi)\Delta U
\end{equation}
At present the tightest upper bound for $\xi$ is of order $10^{-4}$ \cite{vessot1979test,vessot1980test}\footnote{A new, more strict limit has recently been proposed which could push this limit to $10^{-7}$ \cite{Muller:2010zz} or even $10^{-9}$. The validity of this method is, however, still matter of debate \cite{Wolf:2010na,Muller:2010pb}. In our calculations we will not consider this new bound because, as we demonstrate, even such an improvement will not return a significant bound for $a$.}.

A central point in Eq.~\eqref{z-original} is that one will obtain $\xi=0$ in accordance with the equivalence principle if one considers the difference in effective potential \eqref{eq:rind11} with $\ell=0$. In what follows, however, we assume that the deviation induced by $\xi$ is due to the Rindler acceleration, so that $\xi=\xi(a)$. In other words we assume that  the corrections to the gravitational redshift formula are due to our ignorance of the actual  spacetime geometry. Using \eqref{z-original} together with \eqref{RedshiftNewt} yields
\begin{equation}\label{redshiftrind}
a=\frac{\xi M}{r_E r_R}\ .
\end{equation}
The result \eqref{redshiftrind} connects directly the value of $\xi$ with the Rindler acceleration $a$ and allows, once the other quantities are known, to convert the limits on $\xi$ into limits of $a$. This conversion is, unfortunately, not always possible due to the lack of information on these quantities in the literature. For this reason we will restrict ourselves to a single representative example. 

Let us  consider the case of the redshift of the light of the Sun arriving to Jupiter following a study based on data coming from the Galileo probe \cite{krisher1993galileo}. We obtain
\begin{equation}
a\lesssim 5 \cdot 10^{-57}\sim 3 \cdot 10^{-5}m/s^{2}
\end{equation}
which is a much weaker constraint than the ones derived from perihelion shifts or radar echo delay. 
We conclude that gravitational redshift is not  a crucial testing ground for the Rindler acceleration.

So far we have considered only weak field tests.
Strong field tests are not expected to yield relevant bounds on Rindler acceleration, since strong fields imply small radial distances where Rindler acceleration becomes negligible.
To demonstrate this by a simple example we calculate now the shift of the innermost stable circular orbit (ISCO) of a test-particle around a central object with mass $M$ and Rindler constant $a$.
An appreciable shift of the radius of the ISCO would have important consequences for the inner dynamics of accretion disks, see for instance Ref.~\cite{Abramowicz:2008bk}.
We expand in the small parameter $aM$ and obtain by virtue of $\extd V^{\rm eff}/\extd r=\extd^2V^{\rm eff}/\extd r^2=0$ from \eqref{eq:rind11}
\eq{
r_{\textrm{\tiny ISCO}} = 6M\,\big(1-36aM\big) + \dots\qquad \ell_{\textrm{\tiny ISCO}} = \pm2\sqrt{3}M\,\big(1+12aM\big)+\dots\ .
}{eq:ct14}
Similarly, the position of the unstable circular orbit for light like particles follows from \eqref{eq:ct1}, setting $\extd V^{\rm eff}/\extd r=0$:
\eq{
r_{\textrm{\tiny null}} = 3M\,\big(1-3aM\big)+\dots
}{eq:ct15}
The next-to-leading order terms are suppressed by ${\cal O}(aM)$ as compared to the leading order terms.
Inserting the weak upper bound \eqref{eq:Pioneerbound} for $a$ and one solar mass for $M$ yields a tiny number, $aM\approx 10^{-23}$.
This means that for a solar size black hole the ISCO would change at most by a radial distance of $10^{-20}m$.
Even for the biggest black hole observed so far, OJ287 with 18 billion solar masses, and the biggest Rindler acceleration compatible with the Pioneer data, Eq.~\eqref{eq:Pioneerbound}, we obtain a relative suppression by $10^{-13}$ in the correction to the ISCO, which amounts to a shift of approximately $1m$ only.
Thus, the effects from Rindler acceleration are negligible for the location of the ISCO. 
We expect similar conclusions for other strong field tests.

\section{Discussion and outlook}\label{se:5}

In summary, the tightest bound on Rindler acceleration we found, Eq.~\eqref{eq:ct6}, was derived from the data of the Mars and Earth perihelion shifts. 
A bound that is stronger by about an order of magnitude was derived recently by Iorio \cite{Iorio:2010tp} from the Mars range and range-range residuals, $|a| < 1.8\cdot 10^{-67} \sim 10^{-15} m/s^2$.
This means that a universal solar system Rindler acceleration acting on Earth or Mars must be at least five-six orders of magnitude smaller than the Pioneer acceleration in Eq.~\eqref{eq:Pioneerbound}.
We conclude that the Pioneer anomaly cannot be caused by a universal solar system Rindler acceleration. This conclusion is compatible with the analysis by Exirifard \cite{Exirifard:2009zz}, who studied higher curvature theories of gravity and concluded that they cannot simultaneously explain the Pioneer anomaly and be compatible with solar system precision data.

Here some cautionary remarks are in order. Since the assumption of spherical symmetry is crucial in the derivation of the Rindler acceleration in Ref.~\cite{Grumiller:2010bz}, we have to analyze if the effects from breaking of spherical symmetry by test-objects like planets are sufficiently small. Such an analysis is complicated by the fact that so far there is no microscopic understanding of the emergence of a Rindler force. The best we can do is to provide a heuristic bound. We denote solar mass, radial distance, test-object density and test-object radius by $M$, $r$, $\rho$ and $r_0$, respectively.
An obvious requirement is $M\gg \rho r_0^3$, which certainly holds for all solar system objects, including planets.
We treat now the test-object as consisting of concentric spherical shells and consider the energy budget in the outermost shell.
The self-gravity of the test-object leads to a potential energy (per unit mass) of the order $\rho r_0^2$.
The energy (per unit mass) coming from the Rindler force is given by $ar$. If the former is considerably larger than the latter, it is not justified for the outermost shell to consider only the gravitational field (including Rindler force) of the Sun. Instead, a 2-body problem must be considered, which breaks spherical symmetry. It is not clear to what extent Rindler acceleration would emerge in such a 2-body problem. In fact, it appears plausible that the Rindler acceleration takes its maximal value when the test-particle has a negligible mass and tends to zero for symmetry reasons if the ``test-particle'' approaches the mass of the central object. Then, for any test-object there is a critical radius beyond which one should not assume the universal (=maximal) value for the Rindler acceleration, but rather a (possibly considerably) smaller value. We estimate now this critical radius $r_0^{\rm crit}$ by equating the gravitational self-energy of the object with the Rindler energy.
\eq{
r_0^{\rm crit} \sim \sqrt{\frac{ar}{\rho}}
}{eq:angelinajolie}
In the following order-of-magnitude estimates we always use the maximal value $a\approx 10^{-61}$ for the Rindler constant.
For Earth we have $r\approx 10^{46}$ and $\rho\approx10^{-93}$. 
This yields $r_0^{\rm crit} \approx 10^{39}\sim 16km$.
Thus, any object at Earth's orbit with density comparable to Earth's density should not be much larger than a few kilometers in order to avoid a strong violation of the inequality $r_0 < r_0^{\rm crit}$.
Clearly, Earth violates this bound strongly.
On the other hand, the Pioneer spacecraft is considerably smaller than a kilometer, and thus satisfies the bound $r_0 < r_0^{\rm crit}$~\footnote{%
Intriguingly, in the system Galaxy/Sun the radius $r_0$ is close to the critical radius $r_0^{\rm crit}$. 
This observation may be of relevance for interpreting galactic rotation curves in the presence of a Rindler force.}. For the near Earth object (1566) Icarus with a diameter of about $1km$, see table \ref{tab:1}, the bound is also (marginally) satisfied. Certainly, the conclusions from the heuristic discussion above only affect massive test-objects, and thus the only bounds derived in our paper that are potentially weakened by it are the ones from perihelion shifts in section II. The bounds extracted from geodesics of light rays, in particular the tightest bound \eqref{eq:ct13}, remain intact. We can therefore state with confidence that a universal solar system Rindler acceleration should be smaller than three nanometers per second per second. Future space missions to outer planets (e.g. Juno \cite{Juno} or
EJSM-Laplace \cite{Laplace}) will provide further opportunities for precision measurements of the radar
echo delay and might improve the bound \eqref{eq:ct13}.

It would be of great interest to derive the heuristic bound \eqref{eq:angelinajolie} --- or to replace it by a different bound --- from first principles.
Furthermore, it would be interesting to drive the tests for the Rindler acceleration to larger distance scales, i.e. galactic and cosmological scales. For example one could consider gravitational lensing for spherical galaxies, the influence of the Rindler accleration of primordial black holes on the dynamics of the early universe and Rindler corrections to the late time evolution of the universe. In all the mentioned aspects it would be necessary to develop a cosmological model including the Rindler acceleration. We leave this to future work.

\acknowledgments

We thank Lorenzo Iorio for correspondence and Leopold Summerer for encouragement. We also thank Herbert Balasin for discussions.

DG was supported by the START project Y435-N16 of the Austrian Science Foundation (FWF).
FP was supported by the FWF project P22114-N16.
DG and FP acknowledge support from the ESA Advanced Concept Team project Ariadna ID 07/1301, AO/1-5582/07/NL/CB.


\begin{thebibliography}{34}
\expandafter\ifx\csname natexlab\endcsname\relax\def\natexlab#1{#1}\fi
\expandafter\ifx\csname bibnamefont\endcsname\relax
  \def\bibnamefont#1{#1}\fi
\expandafter\ifx\csname bibfnamefont\endcsname\relax
  \def\bibfnamefont#1{#1}\fi
\expandafter\ifx\csname citenamefont\endcsname\relax
  \def\citenamefont#1{#1}\fi
\expandafter\ifx\csname url\endcsname\relax
  \def\url#1{\texttt{#1}}\fi
\expandafter\ifx\csname urlprefix\endcsname\relax\def\urlprefix{URL }\fi
\providecommand{\bibinfo}[2]{#2}
\providecommand{\eprint}[2][]{\url{#2}}

\bibitem[{\citenamefont{Grumiller}(2010)}]{Grumiller:2010bz}
\bibinfo{author}{\bibfnamefont{D.}~\bibnamefont{Grumiller}},
  \bibinfo{journal}{Phys.Rev.Lett.} \textbf{\bibinfo{volume}{105}},
  \bibinfo{pages}{211303} (\bibinfo{year}{2010}), \eprint{1011.3625,
  Erratum-ibid. {\bf 106} (2010), 039901}.

\bibitem[{\citenamefont{Wald}(1984)}]{waldgeneral}
\bibinfo{author}{\bibfnamefont{R.~M.} \bibnamefont{Wald}},
  \emph{\bibinfo{title}{{General Relativity}}} (\bibinfo{publisher}{The
  University of Chicago Press}, \bibinfo{year}{1984}).

\bibitem[{\citenamefont{Anderson et~al.}(1998)}]{Anderson:1998jd}
\bibinfo{author}{\bibfnamefont{J.~D.} \bibnamefont{Anderson}}
  \bibnamefont{et~al.}, \bibinfo{journal}{Phys. Rev. Lett.}
  \textbf{\bibinfo{volume}{81}}, \bibinfo{pages}{2858} (\bibinfo{year}{1998}),
  \eprint{gr-qc/9808081}.

\bibitem[{\citenamefont{Weinberg}(1972)}]{Weinberg:1972}
\bibinfo{author}{\bibfnamefont{S.}~\bibnamefont{Weinberg}},
  \emph{\bibinfo{title}{Gravitation and cosmology: principles and applications
  of the general theory of relativity}} (\bibinfo{publisher}{Wiley},
  \bibinfo{address}{New York}, \bibinfo{year}{1972}).

\bibitem[{\citenamefont{Pitjeva}(2005)}]{Pitjeva:2005}
\bibinfo{author}{\bibfnamefont{E.~V.} \bibnamefont{Pitjeva}},
  \bibinfo{journal}{Astron. Lett.} \textbf{\bibinfo{volume}{31}},
  \bibinfo{pages}{340 } (\bibinfo{year}{2005}).

\bibitem[{\citenamefont{Iorio}(2008)}]{Iorio:2007ub}
\bibinfo{author}{\bibfnamefont{L.}~\bibnamefont{Iorio}},
  \bibinfo{journal}{Advances in Astronomy} \textbf{\bibinfo{volume}{2008}},
  \bibinfo{pages}{Article ID 268647} (\bibinfo{year}{2008}),
  \eprint{0710.2610}.

\bibitem[{\citenamefont{Sereno and Jetzer}(2006)}]{Sereno:2006mw}
\bibinfo{author}{\bibfnamefont{M.}~\bibnamefont{Sereno}} \bibnamefont{and}
  \bibinfo{author}{\bibfnamefont{P.}~\bibnamefont{Jetzer}},
  \bibinfo{journal}{Mon. Not. Roy. Astron. Soc.}
  \textbf{\bibinfo{volume}{371}}, \bibinfo{pages}{626} (\bibinfo{year}{2006}),
  \eprint{astro-ph/0606197}.

\bibitem[{\citenamefont{Milgrom}(1983)}]{Milgrom:1983ca}
\bibinfo{author}{\bibfnamefont{M.}~\bibnamefont{Milgrom}},
  \bibinfo{journal}{Astrophys. J.} \textbf{\bibinfo{volume}{270}},
  \bibinfo{pages}{365} (\bibinfo{year}{1983}).

\bibitem[{\citenamefont{Shapiro et~al.}(2004)\citenamefont{Shapiro, Davis,
  Lebach, and Gregory}}]{Shapiro:2003}
\bibinfo{author}{\bibfnamefont{S.~S.} \bibnamefont{Shapiro}},
  \bibinfo{author}{\bibfnamefont{J.~L.} \bibnamefont{Davis}},
  \bibinfo{author}{\bibfnamefont{D.~E.} \bibnamefont{Lebach}},
  \bibnamefont{and} \bibinfo{author}{\bibfnamefont{J.~S.}
  \bibnamefont{Gregory}}, \bibinfo{journal}{Phys. Rev. Lett.}
  \textbf{\bibinfo{volume}{92}}, \bibinfo{pages}{121101}
  (\bibinfo{year}{2004}).

\bibitem[{\citenamefont{Bertotti et~al.}(2003)\citenamefont{Bertotti, Iess, and
  Tortora}}]{Bertotti:2003}
\bibinfo{author}{\bibfnamefont{B.}~\bibnamefont{Bertotti}},
  \bibinfo{author}{\bibfnamefont{L.}~\bibnamefont{Iess}}, \bibnamefont{and}
  \bibinfo{author}{\bibfnamefont{P.}~\bibnamefont{Tortora}},
  \bibinfo{journal}{Nature (London)} \textbf{\bibinfo{volume}{475}},
  \bibinfo{pages}{374} (\bibinfo{year}{2003}).

\bibitem[{\citenamefont{Will}(2010)}]{will2010confrontation}
\bibinfo{author}{\bibfnamefont{C.}~\bibnamefont{Will}},
  \bibinfo{journal}{General Relativity and John Archibald Wheeler} pp.
  \bibinfo{pages}{73--93} (\bibinfo{year}{2010}).

\bibitem[{\citenamefont{Pound and Rebka~Jr}(1960)}]{pound1960apparent}
\bibinfo{author}{\bibfnamefont{R.}~\bibnamefont{Pound}} \bibnamefont{and}
  \bibinfo{author}{\bibfnamefont{G.}~\bibnamefont{Rebka~Jr}},
  \bibinfo{journal}{Physical Review Letters} \textbf{\bibinfo{volume}{4}},
  \bibinfo{pages}{337} (\bibinfo{year}{1960}), ISSN \bibinfo{issn}{1079-7114}.

\bibitem[{\citenamefont{Hafele and
  Keating}(1972{\natexlab{a}})}]{hafele1972around}
\bibinfo{author}{\bibfnamefont{J.}~\bibnamefont{Hafele}} \bibnamefont{and}
  \bibinfo{author}{\bibfnamefont{R.}~\bibnamefont{Keating}},
  \bibinfo{journal}{Science} \textbf{\bibinfo{volume}{177}},
  \bibinfo{pages}{168} (\bibinfo{year}{1972}{\natexlab{a}}).

\bibitem[{\citenamefont{Hafele and
  Keating}(1972{\natexlab{b}})}]{hafele1972-obs}
\bibinfo{author}{\bibfnamefont{J.}~\bibnamefont{Hafele}} \bibnamefont{and}
  \bibinfo{author}{\bibfnamefont{R.}~\bibnamefont{Keating}},
  \bibinfo{journal}{Science} \textbf{\bibinfo{volume}{177}},
  \bibinfo{pages}{166} (\bibinfo{year}{1972}{\natexlab{b}}).

\bibitem[{\citenamefont{Snider}(1972)}]{snider1972new}
\bibinfo{author}{\bibfnamefont{J.}~\bibnamefont{Snider}},
  \bibinfo{journal}{Physical Review Letters} \textbf{\bibinfo{volume}{28}},
  \bibinfo{pages}{853} (\bibinfo{year}{1972}).

\bibitem[{\citenamefont{Colella et~al.}(1975)\citenamefont{Colella, Overhauser,
  and Werner}}]{Colella:1975dq}
\bibinfo{author}{\bibfnamefont{R.}~\bibnamefont{Colella}},
  \bibinfo{author}{\bibfnamefont{A.}~\bibnamefont{Overhauser}},
  \bibnamefont{and} \bibinfo{author}{\bibfnamefont{S.}~\bibnamefont{Werner}},
  \bibinfo{journal}{Phys.Rev.Lett.} \textbf{\bibinfo{volume}{34}},
  \bibinfo{pages}{1472} (\bibinfo{year}{1975}).

\bibitem[{\citenamefont{Vessot and Levine}(1979)}]{vessot1979test}
\bibinfo{author}{\bibfnamefont{R.}~\bibnamefont{Vessot}} \bibnamefont{and}
  \bibinfo{author}{\bibfnamefont{M.}~\bibnamefont{Levine}},
  \bibinfo{journal}{General relativity and gravitation}
  \textbf{\bibinfo{volume}{10}}, \bibinfo{pages}{181} (\bibinfo{year}{1979}),
  ISSN \bibinfo{issn}{0001-7701}.

\bibitem[{\citenamefont{Vessot et~al.}(1980)\citenamefont{Vessot, Levine,
  Mattison, Blomberg, Hoffman, Nystrom, Farrel, Decher, Eby, Baugher
  et~al.}}]{vessot1980test}
\bibinfo{author}{\bibfnamefont{R.}~\bibnamefont{Vessot}},
  \bibinfo{author}{\bibfnamefont{M.}~\bibnamefont{Levine}},
  \bibinfo{author}{\bibfnamefont{E.}~\bibnamefont{Mattison}},
  \bibinfo{author}{\bibfnamefont{E.}~\bibnamefont{Blomberg}},
  \bibinfo{author}{\bibfnamefont{T.}~\bibnamefont{Hoffman}},
  \bibinfo{author}{\bibfnamefont{G.}~\bibnamefont{Nystrom}},
  \bibinfo{author}{\bibfnamefont{B.}~\bibnamefont{Farrel}},
  \bibinfo{author}{\bibfnamefont{R.}~\bibnamefont{Decher}},
  \bibinfo{author}{\bibfnamefont{P.}~\bibnamefont{Eby}},
  \bibinfo{author}{\bibfnamefont{C.}~\bibnamefont{Baugher}},
  \bibnamefont{et~al.}, \bibinfo{journal}{Physical Review Letters}
  \textbf{\bibinfo{volume}{45}}, \bibinfo{pages}{2081} (\bibinfo{year}{1980}),
  ISSN \bibinfo{issn}{1079-7114}.

\bibitem[{\citenamefont{Turneaure et~al.}(1983)\citenamefont{Turneaure, Will,
  Farrell, Mattison, and Vessot}}]{Turneaure:1983zz}
\bibinfo{author}{\bibfnamefont{J.~P.} \bibnamefont{Turneaure}},
  \bibinfo{author}{\bibfnamefont{C.~M.} \bibnamefont{Will}},
  \bibinfo{author}{\bibfnamefont{B.~F.} \bibnamefont{Farrell}},
  \bibinfo{author}{\bibfnamefont{E.~M.} \bibnamefont{Mattison}},
  \bibnamefont{and} \bibinfo{author}{\bibfnamefont{R.~F.~C.}
  \bibnamefont{Vessot}}, \bibinfo{journal}{Phys.Rev.}
  \textbf{\bibinfo{volume}{D27}}, \bibinfo{pages}{1705} (\bibinfo{year}{1983}).

\bibitem[{\citenamefont{Bonse and Wroblewski}(1984)}]{Bonse:1984td}
\bibinfo{author}{\bibfnamefont{U.}~\bibnamefont{Bonse}} \bibnamefont{and}
  \bibinfo{author}{\bibfnamefont{T.}~\bibnamefont{Wroblewski}},
  \bibinfo{journal}{Phys.Rev.} \textbf{\bibinfo{volume}{D30}},
  \bibinfo{pages}{1214} (\bibinfo{year}{1984}).

\bibitem[{\citenamefont{Vucetich et~al.}(1988)}]{Vucetich:1988uw}
\bibinfo{author}{\bibfnamefont{H.}~\bibnamefont{Vucetich}}
  \bibnamefont{et~al.}, \bibinfo{journal}{Phys. Rev.}
  \textbf{\bibinfo{volume}{D38}}, \bibinfo{pages}{2930} (\bibinfo{year}{1988}).

\bibitem[{\citenamefont{Horvath et~al.}(1988)\citenamefont{Horvath, Logiudice,
  Riveros, and Vucetich}}]{Horvath:1988vf}
\bibinfo{author}{\bibfnamefont{J.~E.}~\bibnamefont{Horvath}},
  \bibinfo{author}{\bibfnamefont{E.~A.}~\bibnamefont{Logiudice}},
  \bibinfo{author}{\bibfnamefont{C.}~\bibnamefont{Riveros}}, \bibnamefont{and}
  \bibinfo{author}{\bibfnamefont{H.}~\bibnamefont{Vucetich}},
  \bibinfo{journal}{Phys.Rev.} \textbf{\bibinfo{volume}{D38}},
  \bibinfo{pages}{1754} (\bibinfo{year}{1988}).

\bibitem[{\citenamefont{Krisher et~al.}(1990)\citenamefont{Krisher, Anderson,
  and Campbell}}]{Krisher:1990zz}
\bibinfo{author}{\bibfnamefont{T.~P.} \bibnamefont{Krisher}},
  \bibinfo{author}{\bibfnamefont{J.~D.} \bibnamefont{Anderson}},
  \bibnamefont{and} \bibinfo{author}{\bibfnamefont{J.~K.}
  \bibnamefont{Campbell}}, \bibinfo{journal}{Phys. Rev. Lett.}
  \textbf{\bibinfo{volume}{64}}, \bibinfo{pages}{1322} (\bibinfo{year}{1990}).

\bibitem[{\citenamefont{Krisher et~al.}(1993)\citenamefont{Krisher, Morabito,
  and Anderson}}]{krisher1993galileo}
\bibinfo{author}{\bibfnamefont{T.}~\bibnamefont{Krisher}},
  \bibinfo{author}{\bibfnamefont{D.}~\bibnamefont{Morabito}}, \bibnamefont{and}
  \bibinfo{author}{\bibfnamefont{J.}~\bibnamefont{Anderson}},
  \bibinfo{journal}{Physical review letters} \textbf{\bibinfo{volume}{70}},
  \bibinfo{pages}{2213} (\bibinfo{year}{1993}), ISSN \bibinfo{issn}{1079-7114}.

\bibitem[{\citenamefont{Godone et~al.}(1995)\citenamefont{Godone, Novero, and
  Tavella}}]{godone1995null}
\bibinfo{author}{\bibfnamefont{A.}~\bibnamefont{Godone}},
  \bibinfo{author}{\bibfnamefont{C.}~\bibnamefont{Novero}}, \bibnamefont{and}
  \bibinfo{author}{\bibfnamefont{P.}~\bibnamefont{Tavella}},
  \bibinfo{journal}{Physical Review D} \textbf{\bibinfo{volume}{51}},
  \bibinfo{pages}{319} (\bibinfo{year}{1995}), ISSN \bibinfo{issn}{1550-2368}.

\bibitem[{\citenamefont{Littrell et~al.}(1997)\citenamefont{Littrell, Allman,
  and Werner}}]{Littrell:1997zz}
\bibinfo{author}{\bibfnamefont{K.~C.}~\bibnamefont{Littrell}},
  \bibinfo{author}{\bibfnamefont{B.~E.}~\bibnamefont{Allman}}, \bibnamefont{and}
  \bibinfo{author}{\bibfnamefont{S.~A.}~\bibnamefont{Werner}},
  \bibinfo{journal}{Phys.Rev.} \textbf{\bibinfo{volume}{A56}},
  \bibinfo{pages}{1767} (\bibinfo{year}{1997}).

\bibitem[{\citenamefont{Cottam et~al.}(2002)\citenamefont{Cottam, Paerels, and
  Mendez}}]{Cottam:2002cu}
\bibinfo{author}{\bibfnamefont{J.}~\bibnamefont{Cottam}},
  \bibinfo{author}{\bibfnamefont{F.}~\bibnamefont{Paerels}}, \bibnamefont{and}
  \bibinfo{author}{\bibfnamefont{M.}~\bibnamefont{Mendez}},
  \bibinfo{journal}{Nature} \textbf{\bibinfo{volume}{420}}, \bibinfo{pages}{51}
  (\bibinfo{year}{2002}), \eprint{astro-ph/0211126}.

\bibitem[{\citenamefont{Cacciapuoti and Salomon}(2009)}]{cacciapuoti2009space}
\bibinfo{author}{\bibfnamefont{L.}~\bibnamefont{Cacciapuoti}} \bibnamefont{and}
  \bibinfo{author}{\bibfnamefont{C.}~\bibnamefont{Salomon}},
  \bibinfo{journal}{The European Physical Journal-Special Topics}
  \textbf{\bibinfo{volume}{172}}, \bibinfo{pages}{57} (\bibinfo{year}{2009}),
  ISSN \bibinfo{issn}{1951-6355}.

\bibitem[{\citenamefont{Haugan}(1979)}]{Haugan:1979iv}
\bibinfo{author}{\bibfnamefont{M.}~\bibnamefont{Haugan}},
  \bibinfo{journal}{Annals Phys.} \textbf{\bibinfo{volume}{118}},
  \bibinfo{pages}{156} (\bibinfo{year}{1979}).

\bibitem[{\citenamefont{Abramowicz}(2008)}]{Abramowicz:2008bk}
\bibinfo{author}{\bibfnamefont{M.~A.} \bibnamefont{Abramowicz}}
  (\bibinfo{year}{2008}), \eprint{0812.3924}.

\bibitem[{\citenamefont{Iorio}(2010)}]{Iorio:2010tp}
\bibinfo{author}{\bibfnamefont{L.}~\bibnamefont{Iorio}} (\bibinfo{year}{2010}),
  \eprint{1012.0226}.

\bibitem[{\citenamefont{Muller et~al.}(2010{\natexlab{a}})\citenamefont{Muller,
  Peters, and Chu}}]{Muller:2010zz}
\bibinfo{author}{\bibfnamefont{H.}~\bibnamefont{Muller}},
  \bibinfo{author}{\bibfnamefont{A.}~\bibnamefont{Peters}}, \bibnamefont{and}
  \bibinfo{author}{\bibfnamefont{S.}~\bibnamefont{Chu}},
  \bibinfo{journal}{Nature} \textbf{\bibinfo{volume}{463}},
  \bibinfo{pages}{926} (\bibinfo{year}{2010}{\natexlab{a}}).

\bibitem[{\citenamefont{Wolf et~al.}(2010)\citenamefont{Wolf, Blanchet, Borde,
  Cohen-Tannoudji, Salomon et~al.}}]{Wolf:2010na}
\bibinfo{author}{\bibfnamefont{P.}~\bibnamefont{Wolf}},
  \bibinfo{author}{\bibfnamefont{L.}~\bibnamefont{Blanchet}},
  \bibinfo{author}{\bibfnamefont{C.~J.} \bibnamefont{Borde}},
  \bibinfo{author}{\bibfnamefont{C.}~\bibnamefont{Cohen-Tannoudji}},
  \bibinfo{author}{\bibfnamefont{C.}~\bibnamefont{Salomon}},
  \bibnamefont{et~al.} (\bibinfo{year}{2010}), \eprint{1012.1194}.

\bibitem[{\citenamefont{Muller et~al.}(2010{\natexlab{b}})\citenamefont{Muller,
  Peters, and Chu}}]{Muller:2010pb}
\bibinfo{author}{\bibfnamefont{H.}~\bibnamefont{Muller}},
  \bibinfo{author}{\bibfnamefont{A.}~\bibnamefont{Peters}}, \bibnamefont{and}
  \bibinfo{author}{\bibfnamefont{S.}~\bibnamefont{Chu}},
  \bibinfo{journal}{Nature} \textbf{\bibinfo{volume}{467}}, \bibinfo{pages}{E2}
  (\bibinfo{year}{2010}{\natexlab{b}}), \eprint{1009.1838}.

\bibitem[{\citenamefont{Exirifard}(2009)}]{Exirifard:2009zz}
\bibinfo{author}{\bibfnamefont{Q.}~\bibnamefont{Exirifard}},
  \bibinfo{journal}{Class.Quant.Grav.} \textbf{\bibinfo{volume}{26}},
  \bibinfo{pages}{025001} (\bibinfo{year}{2009}).

\bibitem{Juno}\bibinfo{author}{\bibfnamefont{B.}~\bibnamefont{Dunbar}},~(\bibinfo{year}{2011}), \eprint{\url{http://www.nasa.gov/mission_pages/juno/main/index.html}}.

\bibitem{Laplace}\bibinfo{author}{\bibnamefont{ESA}},~(\bibinfo{year}{2011}), \eprint{\url{http://sci.esa.int/science-e/www/area/index.cfm?fareaid=107}}.
\end{thebibliography}
\end{document}